\documentclass[prd,superscriptaddress,showpacs,onecolumn]{revtex4-2}
\usepackage{amssymb}

\usepackage{bm}
\usepackage{amsfonts}
\usepackage{latexsym}
\usepackage[latin1]{inputenc}
\usepackage{graphicx}
\usepackage{amsmath}
\usepackage{palatino}
\usepackage{mathpazo}
\usepackage{textcomp}
\usepackage{float}
\usepackage{booktabs}
\usepackage{dcolumn}
\usepackage{hyperref}
\usepackage{amsmath}
\usepackage{xcolor}
\usepackage{orcidlink}
\usepackage[caption=false]{subfig}
\usepackage{commath}

\hypersetup{colorlinks,citecolor=blue}
\def\jnl@style{\it}
\def\aaref@jnl#1{{\jnl@style#1}}
\def\aaref@jnl#1{{\jnl@style#1}}
\def\aj{\aaref@jnl{AJ}}
\def\apj{\aaref@jnl{ApJ}}
\def\apjl{\aaref@jnl{ApJ}}
\def\apjs{\aaref@jnl{ApJS}}
\def\apss{\aaref@jnl{Ap\&SS}}
\def\aap{\aaref@jnl{A\&A}}
\def\aapr{\aaref@jnl{A\&A~Rev.}}
\def\aaps{\aaref@jnl{A\&AS}}
\def\mnras{\aaref@jnl{Mon.~Not.~Roy.~Astron.~Soc.}}
\def\prd{\aaref@jnl{Phys.~Rev.~D}}
\def\prc{\aaref@jnl{Phys.~Rev.~C}}
\def\prl{\aaref@jnl{Phys.~Rev.~Lett.}}
\def\qjras{\aaref@jnl{QJRAS}}
\def\skytel{\aaref@jnl{S\&T}}
\def\ssr{\aaref@jnl{Space~Sci.~Rev.}}
\def\zap{\aaref@jnl{ZAp}}
\def\nat{\aaref@jnl{Nature}}
\def\aplett{\aaref@jnl{Astrophys.~Lett.}}
\def\apspr{\aaref@jnl{Astrophys.~Space~Phys.~Res.}}
\def\physrep{\aaref@jnl{Phys.~Rep.}}
\def\physscr{\aaref@jnl{Phys.~Scr}}
\def\commat{\aaref@jnl{Comm.~Math.~Phys.}}
\def\science{\aaref@jnl{Science}}
\def\cqg{\aaref@jnl{Classical Quant.~Grav.}}
\def\jpcs{\aaref@jnl{JPCS}}
\def\ijmpd{\aaref@jnl{Int.~J.~Mod.~Phys.~D}}
\def\grg{\aaref@jnl{Gen.~Relat.~Gravit.}}
\def\rpp{\aaref@jnl{Rep.~Prog.~Phys.}}
\def\npa{\aaref@jnl{Nucl.~Phys.~A}}
\def\lrr{\aaref@jnl{Living Rev.~Rel.}}
\def\jcap{\aaref@jnl{J.~Cosmology Astropart.~Phys.}}
\def\rmp{\aaref@jnl{Rev.~Mod.~Phys.}}
\def\epjc{\aaref@jnl{Eur.~Phys.~J.~C}}

\allowdisplaybreaks[1]
\renewcommand{\arraystretch}{1.1}
\begin{document}

\title{Cosmological implications of an interacting model of dark matter \&
dark energy}
\author{Keshav Ram Mishra}
\email{krmgkp1995@gmail.com}
\affiliation{Department of Mathematics and Statistics, DDU Gorakhpur
	University, Gorakhpur 273009, Uttar Pradesh, India}

\author{Shibesh Kumar Jas Pacif\orcidlink{0000-0003-0951-414X}}
\email{shibesh.math@gmail.com}
\affiliation{Centre for Cosmology and Science Popularization (CCSP), SGT
	University, Delhi-NCR, Gurugram 122505, Haryana, India.}

\author{Rajesh Kumar\orcidlink{0000-0002-2582-7245}}
\email{rkmath09@gmail.com }
\affiliation{Department of Mathematics and Statistics, DDU Gorakhpur
	University, Gorakhpur 273009, Uttar Pradesh, India}

\author{Kazuharu Bamba\orcidlink{0000-0001-9720-8817}}
\email{bamba@sss.fukushima-u.ac.jp}
\affiliation{Faculty of Symbiotic Systems Science, Fukushima University,
	Fukushima 960-1296, Japan}

\date{\today }

\begin{abstract}
In this paper, we have studied an interacting dark energy model. We have
assumed the gravitational interaction between the matter fields i.e. between
barotropic fluid and the dark energy. The dark energy evolution within the
framework of spatially homogeneous and isotropic Friedmann-Robertson-Walker
space-time. Therefore, we examine the cosmic evolution from the perspective
of interacting scenario by selecting a suitable ansatz for the scale factor
resulting from a parametrization of Hubble parameter. The evolution of the
cosmological parameters are discussed in some details in the considered
interacting scenario by calculating parameters and quantities such as
deceleration parameter, energy density, pressure, equation of state (EoS)
etc. Also, we have performed some cosmological tests and analysis in support
of our obtained interacting model. Finally, we reconstruct the potential of
the scalar field and refute the refined swampland conjecture using the
equation of state of dark energy and the relationship between energy density
and pressure with the scalar field and potential, and then thoroughly
describe the findings.
\end{abstract}

\maketitle

\color{black} 


\section{Introduction}

\label{sec1}

One of the major issues in theoretical physics and more generally, in
cosmology, over the past decades is to determine the enigmatic nature of the
two dominant components in the universe namely, dark energy and dark matter.
This greatest open challenge in cosmology nowadays is the physical reason
underpinning the late-time cosmic speed-up. The physical mechanism has been
revealed by explicate the various statistical observational data sets \cite%
{Reiss1998}, \cite{Perlmutter1999}, \cite{Tegamark2004}, \cite{Hinshaw2013}, 
\cite{Rose2020}, \cite{Jaffe2000}, \cite{CMB2007}, \cite{BAO2006}, \cite%
{SELJAK2004}, \cite{SDSS2006}. Many models of dark energy postulate the
existence of an extra, unknown field that is responsible for the rapid
expansion of the Universe named dark energy, but some viable hypotheses
include infrared modification to the theory of general relativity (for some
reviews, see \cite{ejcope}, \cite{kbam}, \cite{snoj}). Dark energy accounts
for around $68\%$ of the overall energy density of the Universe, and hence
consequently controls the evolution of the current cosmos. \newline

\qquad The cosmological constant $\Lambda $ is the most straightforward dark
energy possibility, with equation of state $(EoS)$ parameter $\omega
_{\Lambda }=\frac{p_{\Lambda }}{\rho _{\Lambda }}=-1$. The $\Lambda $CDM
model sides well with the current cosmological observations \cite{parplanck}%
, and its parameters have been well-established with impressive precision
using latest observational data. On the other hand, $\Lambda $CDM model has
always been beset by obvious theoretical objections such as fine-tuning and
cosmic-coincidence issues \cite{vsahni}, \cite{rbean}. As a result, various
expansions to the underlying $\Lambda $CDM cosmology have received a lot of
attention, including the possibility of vacuum energy interacting with $%
\Lambda $CDM \cite{lamen1}, \cite{lamen2}, \cite{rgcai}, \cite{xzhan}, \cite%
{msas}, \cite{yhli}, \cite{lfeng}, \cite{rmurg}, \cite{xzhan1}, \cite{apou}.
Due to the significant difference between the theoretical and observational
predictions on the value of $\Lambda $ and to alleviate the tension between
the $\Lambda $CDM and the observational data \cite{Weinberg1989} advised to
modify the gravity and move toward the alternating theories of gravity with
an enhancement in the goodness of fit \cite{parplanck1}, \cite{tmc}, \cite%
{nagh}. As a replacement for $\Lambda $CDM model, the rolling scalar field
has been intensively introduced into the literature \cite{pjpee}, \cite%
{evlin}, \cite{mlix}. Resultantly, the significance of seeing the scalar
field as a DE candidate in understanding the unification of cosmic
acceleration (early and late time) has earned high acceptance \cite{ejcope}.%
\newline

\qquad In addition to the shortcomings of the standard model of cosmology,
certain observational results are still puzzling. For instance, in some
studies, $\Lambda $CDM model has been modified by connecting a dynamical
dark energy model with $H_{2.34}$ data. Work in \cite{eaub}, \cite{gbzh}
reveals how to evaluate a range of models that allow for the evolution of
dark energy models based on BAO, CMB, and SN data, as well as how to match
dynamical DE models with $H_{2.34}$ data. In some observations of BOSS data 
\cite{tdel}, the observed value of Hubble parameter $H$\ at redshift $z=2.34$%
\ is $222\pm 7kms^{-1}Mpc^{-1}$\ which falls below the $\Lambda $CDM model's
prediction. \cite{agva}, \cite{rygu} have some additional similar work. The
goal of this study is to create a cosmological model in a manner that
incorporates new phenomenological types of interaction between dark energy
and dark matter. Here, we study the cosmic acceleration in a interacting
scenario between dark matter and dark energy.\textbf{\ }In this sense, some
theories come up with DM and DE interaction, disclosing new properties of
both components.\newline

\qquad Due to a lack of understanding of the nature of the dark sector,
numerous ways have been developed to divulge the physical features of DM and
DE. Searching for models that could provide an alternate means of
investigating the nature of the dark sector and, ideally, allowing for
differentiation between the many theoretical models is still on. In this
sense, some theories propose that dark matter and dark energy interact,
revealing new properties of both (for a recent review see \cite{bwang}).
Indeed, numerous interactions have been proposed, including non-minimally
coupled theories \cite{tkoi, lamen}, where the complete Lagrangian contains
a specific interaction term.\newline

\qquad Dark matter and dark energy interaction is a potential mechanism that
states that in particle physics or more theoretical ground, any two matter
fields can interact with each other. This particular phenomenological theory
has piqued the cosmology community's curiosity due to various conceivable
outcomes. Interacting models of DM and DE are an equivalent description of
the dark sector of the Universe that has been extensively researched and are
motivated by a viable explanation to the so called coincidence and
cosmological constant concerns as in the interaction model dark energy decay
into dark matter \cite{wzim}, \cite{cgbo}, \cite{lpchi}, \cite{femc}, \cite%
{farev}. Some studies make fair assumptions about a DM-DE relationship to
help ease or overcome the conundrum of the coincidence problem. Because the
nature of DE and DM is unknown to us, and interaction is allowed in field
theory \cite{smic}. The energy densities of DE and DM are allied in this
way, and both become dynamical, allowing them to be comparable and reducing
the coincidence. Numerous work \cite{bwang1}, \cite{smic}, \cite{eabda}, 
\cite{eabda1}, \cite{vsal}, \cite{Sunny1}, \cite{Sunny2} have already been
presented in this regard that highlight the relation between DE and DM
interaction. Recently, few interacting models have been explored in the
contexts of resolving some cosmological tensions \cite{Sunny3}, \cite{Sunny4}%
, \cite{Sunny5}, \cite{Interact1}, \cite{Interact2}, \cite{Interact3}, \cite%
{Interact4}, \cite{Interact5}, \cite{Interact7}, \cite{Interact8}, \cite%
{Interact9}, \cite{Interact10}, \cite{Interact11}, \cite{Interact12}, \cite%
{Interact13}, \cite{Interact14}. For a more comprehensive list of references
on interaction evidence discovered so far and the discussion of theoretical
and cosmological features, see \cite{bwang2}. It has been identified that
allowing for interaction can shift the dark energy EoS from quintessence to
phantom, implying that the dark energy EoS parameter is imposed with an
effective quintom type of nature. In phantom scalar field models, phantom
crossing line causes instabilities with negative kinetic correction term. In
addition to this, the interaction theory has been found to be particularly
effective in addressing the Hubble constant $H_{0}$ gap between global and
local measurements.\newline

\qquad Since there is no such overarching rule for recruiting interaction
functions, therefore, a variety of linear and non-linear functions are
presented in the literature and one can select some new functions
apparently. Generally, it is tedious to discuss the dynamics of the
interacting model using a non-linear interaction function, hence these
models are quite uncommon in literature \cite{lpchi}, \cite{fare}, \cite%
{wyang}. Nonetheless, non-linear interacting models are always fascinating
to explore the dynamics of the Universe to see if we can glean any further
information from it. As a result of the motivation by the interacting
scenarios, we examine an interacting model in this paper by adopting a
model-independent approach.\newline

\qquad The work of this paper has been organized as follows: Sect. I
provides a brief introduction to general relativity and current status of
the some hot cosmological problems addressed. In Sect. II, we consider the
spacetime metric and formulated the Einstein. In Sect. III, solution of the
field equations are obtained using a parametrization scheme. In Sect. IV, we
have discussed the energy conditions for our model. We have discussed the
stability of the model through velocity of sound in Sect. V. In Sect. VI, we
have discussed the swampland conjecture and finally we have concluded our
results in Sect. VII.

\section{Field equations in an interacting scenario}

\label{sec2}

We start our analysis will go over the metric given in the form of
homogeneous, isotropic, and spherically symmetric spatially Robertson-Walker
geometry of the form 
\begin{equation}
ds^{2}=-dt^{2}+a(t)^{2}\left[ \frac{dr^{2}}{1-\kappa r^{2}}+r^{2}(d\theta
^{2}+\sin ^{2}\theta d\phi ^{2})\right] .  \label{1}
\end{equation}

Here, $a(t)$ denote the scale factor of the Universe: a function of cosmic
time $t$ and $c=1$. The curvature constant, denoted by the term $\kappa $,
assumes the values $0$, $-1$, $1$ showing flat, open, or closed geometry
respectively.\newline

Let us introduce the Einstein field equations (EFEs) for the Friedmann
Robertson-Walker metric in general theory of relativity as 
\begin{equation}
R_{\mu \nu }-\frac{1}{2}Rg_{\mu \nu }=T_{{\mu \nu }},  \label{2}
\end{equation}%
where LHS of the above equation denotes the geometry of the Universe and RHS
represents the energy momentum tensors of the various components in the
Universe i.e. radiation, baryons, dark matter, and dark energy. The
independent field equations will be,

\begin{equation}
3M_{pl}^{2}\left[ H^{2}+ka^{-2}\right] =\rho _{r}+\rho _{b}+\rho _{m}+\rho
_{d}=\rho _{total}\text{,}  \label{3}
\end{equation}%
\begin{equation}
-M_{pl}^{2}\left[ 2\frac{\ddot{a}}{a}+H^{2}+ka^{-2}\right]
=p_{r}+p_{b}+p_{m}+p_{d}=p_{total}\text{.}  \label{4}
\end{equation}

Here, an overhead dot represents the time derivative and $H=\frac{\dot{a}}{a}
$ is the Hubble parameter. $\rho _{r}$, $\rho _{b}$, $\rho _{m}$, $\rho _{d}$%
, $p_{r}$, $p_{b}$, $p_{m}$, and $p_{d}$ denote the energy densities and
corresponding pressures of various components of the Universe. According to
the Bianchi identity, $G_{ij}^{;j}=0$ leads to $T_{ij}^{;j}=0$, which gives
us the following continuity equation.\newline
\begin{equation}
\dot{\rho}_{total}+3H(\rho _{total}+p_{total})=0\text{.}  \label{5}
\end{equation}

As a result of the discussion of motivation in the introduction, let us
consider the interacting scenario in this article. We consider the
gravitational interaction is between two major dominating components i.e.
dark matter and dark energy, which yields the equations: 
\begin{equation}
\dot{\rho}_{r}+3H(\rho _{r}+p_{r})=0\text{,}  \label{6}
\end{equation}%
\begin{equation}
\dot{\rho}_{b}+3H(\rho _{b}+p_{b})=0\text{,}  \label{7}
\end{equation}%
\begin{equation}
\dot{\rho}_{m}+3H(\rho _{m}+p_{m})=Q(t)\text{,}  \label{8}
\end{equation}%
and 
\begin{equation}
\dot{\rho}_{d}+3H(\rho _{d}+p_{d})=-Q(t)\text{,}  \label{9}
\end{equation}%
where $Q(t)$ represents the energy exchange rate between the two dark
sectors in these equations, with $Q(t)>0$ implying a transfer of energy from
the dark matter to the dark energy sector, and $Q(t)<0$ implying the
opposite. The energy transfer rate between dark and light sector is
determined by the interaction term $Q$. However, the exact nature of it is
still unknown to us. To investigate the topic of dark sector interaction,
certain probable forms of $Q$ should be assumed. In this paper, we look at
the the form 
\begin{equation}
Q=3\gamma H\rho _{m},  \label{10}
\end{equation}%
where $\gamma $ is a coupling constant. The interaction term is exactly
proportional to Hubble parameter $H$ to make the preceding equations (\ref{8}%
), (\ref{9}) to hold the continuity law, since it is needed that the
interaction term be proportional to the inverse unit of time. Therefore, We
have chosen the form of $Q=3\gamma H\rho _{m}$ to reflect this feature.

Let us define the equation of state parameter $\omega $ for various
components of the Universe as $\omega _{i}=\frac{p_{i}}{\rho _{i}}$. Then,
from the field equations (\ref{6}), (\ref{7}), (\ref{8}) and (\ref{9})
together (\ref{10}) yield the solution, 
\begin{equation}
\rho _{r}=C_{1}a^{-4}\text{,}  \label{10a}
\end{equation}%
\begin{equation}
\rho _{b}=C_{2}a^{-3}\text{,}  \label{10b}
\end{equation}%
\begin{equation}
\rho _{m}=C_{3}a^{3\gamma -3}\text{,}  \label{11}
\end{equation}%
and%
\begin{equation}
\dot{\rho}_{de}+3H\left( 1+\omega _{de}\right) \rho _{de}=-3C_{3}\gamma
Ha^{3\gamma -3}\text{,}  \label{12}
\end{equation}%
where $C_{1}$, $C_{2}$ and $C_{3}$ are integrating constants. Let us now
define the redshift $z=\frac{a_{0}}{a}-1$. We use the standard lore and
normalize scale factor $a_{0}=1$, henceforth. Equations (\ref{10a}), (\ref%
{10b}) and (\ref{11}) can now be written as,

\begin{equation}
\rho _{r}=C_{1}\left( 1+z\right) ^{4}\text{, }\rho _{b}=C_{2}\left(
1+z\right) ^{3}\text{, }\rho _{m}=C_{3}\left( 1+z\right) ^{3-3\gamma }\text{,%
}  \label{12a}
\end{equation}%
and the expressions of $\rho _{de}$ and $p_{de}$ can be derived in terms of
geometrical parameters as, 
\begin{equation}
p_{d}=M_{pl}^{2}\left[ \left( 2q-1\right) H^{2}-k\left( 1+z\right) ^{2}%
\right] -\frac{1}{3}C_{1}\left( 1+z\right) ^{4}\text{,}  \label{13}
\end{equation}

\begin{equation}
\rho _{d}=3M_{pl}^{2}\left[ H^{2}+k\left( 1+z\right) ^{2}\right] -\left[
C_{1}\left( 1+z\right) ^{4}+C_{2}\left( 1+z\right) ^{3}+C_{3}\left(
1+z\right) ^{3-3\gamma }\right] \text{.}  \label{14}
\end{equation}

Let us now define the density parameter ($\Omega $) as $\Omega _{i}=\frac{%
\rho _{i}}{\rho _{c}}$, where $\rho _{c}=3M_{pl}^{2}H^{2}$ stands for the
critical density and the suffix $i=r$, $b$, $m$ (radiation, baryon and dark
matter). Then, we have

\begin{equation}
\Omega _{r}=\Omega _{r0}\left( 1+z\right) ^{4}\left( \frac{H_{0}}{H}\right)
^{2}\text{, }\Omega _{b}=\Omega _{b0}\left( 1+z\right) ^{3}\left( \frac{H_{0}%
}{H}\right) ^{2}\text{, }\Omega _{m}=\Omega _{m0}\left( 1+z\right)
^{3-3\gamma }\left( \frac{H_{0}}{H}\right) ^{2}\text{.}  \label{14a}
\end{equation}%
Here, the suffix $0$ stands for the values of the cosmological parameters at
present time ($t=t_{0}$ or $z=0$). The friedmann equation (\ref{3}) can be
expressed in terms of density parameter as

\begin{equation}
\Omega _{d}=\left( 1+\Omega _{k}\right) -\left( \frac{H_{0}}{H}\right) ^{2}%
\left[ \Omega _{r0}\left( 1+z\right) ^{4}+\Omega _{b0}\left( 1+z\right)
^{3}+\Omega _{m0}\left( 1+z\right) ^{3-3\gamma }\right] \text{,}  \label{14b}
\end{equation}%
with the understanding of $\frac{k\left( 1+z\right) ^{2}}{H^{2}}=\frac{k}{%
a^{2}H^{2}}=\Omega _{k}$. Finally, we have 
\begin{equation}
p_{d}=M_{pl}^{2}\left[ \left( 2q-1\right) H^{2}-k\left( 1+z\right) ^{2}%
\right] -H_{0}^{2}\Omega _{r0}\left( 1+z\right) ^{4}\text{,}  \label{14c}
\end{equation}

\begin{equation}
\rho _{de}=3M_{pl}^{2}\left[ H^{2}+k\left( 1+z\right) ^{2}\right] -H_{0}^{2}%
\left[ \Omega _{r0}\left( 1+z\right) ^{4}+\Omega _{b0}\left( 1+z\right)
^{3}+\Omega _{m0}\left( 1+z\right) ^{3-3\gamma }\right] \text{,}  \label{14d}
\end{equation}%
and

\begin{equation}
\omega _{d}=\frac{1}{3}\frac{\left( 2q-1\right) H^{2}-k\left( 1+z\right)
^{2}-H_{0}^{2}\Omega _{r0}\left( 1+z\right) ^{4}}{H^{2}+k\left( 1+z\right)
^{2}-H_{0}^{2}\left[ \Omega _{r0}\left( 1+z\right) ^{4}+\Omega _{b0}\left(
1+z\right) ^{3}+\Omega _{m0}\left( 1+z\right) ^{3-3\gamma }\right] }
\label{14e}
\end{equation}

From the calculations above, it is now clear that we can have a fully
deterministic solution of the field equations for any known $H(z)$ form and
then we can explain the characteristics of the model. In order to close the
above system of equations, we need one more equation. In literature, there
are several phenomenological arguments to choose this extra constrain
equation. One of the simplest way is the model-independent way approach that
considers a functional form of any physical or geometrical parameter
involving a few unknown constants (or model parameters). The
model-independent way is also called the parametrization scheme that do not
violate the background physics and provide a simple method to find a general
solution to the system of equations. There is a huge list of various
parametrization schemes summarized in a paper \cite{SKJP-EPJP} those are
used to solve the system of equations in general relativity or in modified
theories of gravity. In the studies of cosmological modelling, this
cosmological parametrization scheme serves as a natural way to discuss the
cosmological dynamics of the model or reconstructing the cosmic evolution.
With this motivation, here we proceed our study of interacting model of dark
matter - dark energy by considering a well-motivated parameterization of
Hubble parameter ($H$) that already considered in several research papers in
different contexts works by Singh \cite{JPS}, Banerjee et al. \cite{NBSD},
Nagpal et al. \cite{RNSKJP}, Pacif et al. \cite{SKJPMSK}, and Mandal et al. 
\cite{SMSBSKJP}: 
\begin{equation}
H(a)=\alpha (1+a^{-n}),  \label{15}
\end{equation}%
where $\alpha >0$ and $n>1$ are constants (call them as model parameters),
which is to be constrained through some observational datasets. In the next
section, we shall discuss the dynamics of the model with the considered
cosmological parametrization.

\section{The Model}

\label{sec3}

The form of Hubble parameter considered in equation (\ref{15}) provides a
smooth dynamics of expansion of the Universe from a decelerating phase to an
accelerating one \cite{RNSKJP}. Equation (\ref{15}) readily yields, the
explicit form of scale factor as,

\begin{equation}
a(t)=(e^{n\alpha t}-1)^{\frac{1}{n}}+c\text{.}  \label{16}
\end{equation}

Using the initial big bang condition (at $t=0$, $a=0$), which make the
constant of integration $c$, zero. We can establish the $t-z$ relationship
as $t(z)=\frac{1}{n\alpha }\log \left( 1+(1+z)^{-n}\right) $ and we can
write, 
\begin{equation}
H(z)=\alpha (1+(1+z)^{n}),  \label{17}
\end{equation}%
or 
\begin{equation}
H(z)=\frac{H_{0}}{2}(1+(1+z)^{n})\text{.}  \label{18}
\end{equation}%
The constant $\alpha $ is of the order of $H_{0}$ and $\alpha =\frac{H_{0}}{2%
}$. The deceleration parameter $q(z)$ comes out to be

\begin{equation}
q(z)=\frac{\left( n-1\right) (1+z)^{n}-1}{1+(1+z)^{n}}  \label{19}
\end{equation}

The detailed dynamical behavior of these geometrical parameters are
discussed in Nagpal et al. \cite{RNSKJP}, Pacif et al. \cite{SKJPMSK}, and
Mandal et al. \cite{SMSBSKJP} in different scenarios in classical and in a
modified theory. Here, in this paper, we are trying to discuss the physical
dynamics of an interacting model of dark energy with this parametrization.
We can observe from equations (\ref{18}) and (\ref{19}) that the expressions
contain only one model parameter $n$. A suitable value of the $n$ would
provide the evolution of different cosmological parameters in our model,
which can be obtained by constraining it with any cosmological datasets. In
the references \cite{RNSKJP} and \cite{SMSBSKJP}, the authors have found the
constrained values of $n=1.43$, to which we are going to use here for our
further analysis. Together with equations (\ref{18}) and (\ref{19}), we
obtain the physical parameters as follows: 
\begin{equation}
\Omega _{r}=\Omega _{r0}\frac{4(1+z)^{4}}{(1+(1+z)^{n})^{2}}\text{,}
\label{20}
\end{equation}%
\begin{equation}
\Omega _{b}=\Omega _{b0}\frac{4(1+z)^{3}}{(1+(1+z)^{n})^{2}}\text{,}
\label{21}
\end{equation}%
\begin{equation}
\Omega _{m}=\Omega _{m0}\frac{4(1+z)^{-3(\gamma -1)}}{(1+(1+z)^{n})^{2}}%
\text{,}  \label{22}
\end{equation}

\begin{equation}
\Omega _{d}=\left( 1+\Omega _{k}\right) -\frac{4\left[ \Omega _{r0}\left(
1+z\right) ^{4}+\Omega _{b0}\left( 1+z\right) ^{3}+\Omega _{m0}\left(
1+z\right) ^{3-3\gamma }\right] }{(1+(1+z)^{n})^{2}}\text{.}  \label{23}
\end{equation}%
Equation (\ref{23}) is verified with $\Omega _{d0}=1+\Omega _{k}-(\Omega
_{r0}+\Omega _{b0}+\Omega _{m0})$. Equations (\ref{14c}), (\ref{14d}) and (%
\ref{14e}) can be written with the help of equations (\ref{18}) and (\ref{19}%
) as, 
\begin{equation}
\frac{p_{d}}{M_{pl}^{2}H_{0}^{2}}=\left[ \frac{\left\{ \left( 2n-3\right)
\left( 1+z\right) ^{n}-3\right\} \left\{ 1+\left( 1+z\right) ^{n}\right\} }{4%
}-\frac{\Omega _{k}\left\{ 1+\left( 1+z\right) ^{n}\right\} ^{2}}{4}-\Omega
_{r0}\left( 1+z\right) ^{4}\right] \text{,}  \label{24}
\end{equation}

\begin{equation}
\frac{\rho _{d}}{3M_{pl}^{2}H_{0}^{2}}=\left[ \frac{\left\{ 1+\left(
1+z\right) ^{n}\right\} ^{2}}{4}+\frac{\Omega _{k}\left\{ 1+\left(
1+z\right) ^{n}\right\} ^{2}}{4}-\left\{ \Omega _{r0}\left( 1+z\right)
^{4}+\Omega _{b0}\left( 1+z\right) ^{3}+\Omega _{m0}\left( 1+z\right)
^{3-3\gamma }\right\} \right] \text{,}  \label{25}
\end{equation}%
and

\begin{equation}
\omega _{d}=\frac{1}{3}\frac{\left\{ \left( 2n-3\right) \left( 1+z\right)
^{n}-3\right\} \left\{ 1+\left( 1+z\right) ^{n}\right\} -\Omega _{k}\left\{
1+\left( 1+z\right) ^{n}\right\} ^{2}-4\Omega _{r0}\left( 1+z\right) ^{4}}{%
\left\{ 1+\left( 1+z\right) ^{n}\right\} ^{2}+\Omega _{k}\left\{ 1+\left(
1+z\right) ^{n}\right\} ^{2}-4\left\{ \Omega _{r0}\left( 1+z\right)
^{4}+\Omega _{b0}\left( 1+z\right) ^{3}+\Omega _{m0}\left( 1+z\right)
^{3-3\gamma }\right\} }  \label{26}
\end{equation}

\begin{figure}[]
\begin{center}
$%
\begin{array}{c@{\hspace{.1in}}cc}
\includegraphics[width=2.2 in, height=2.2 in]{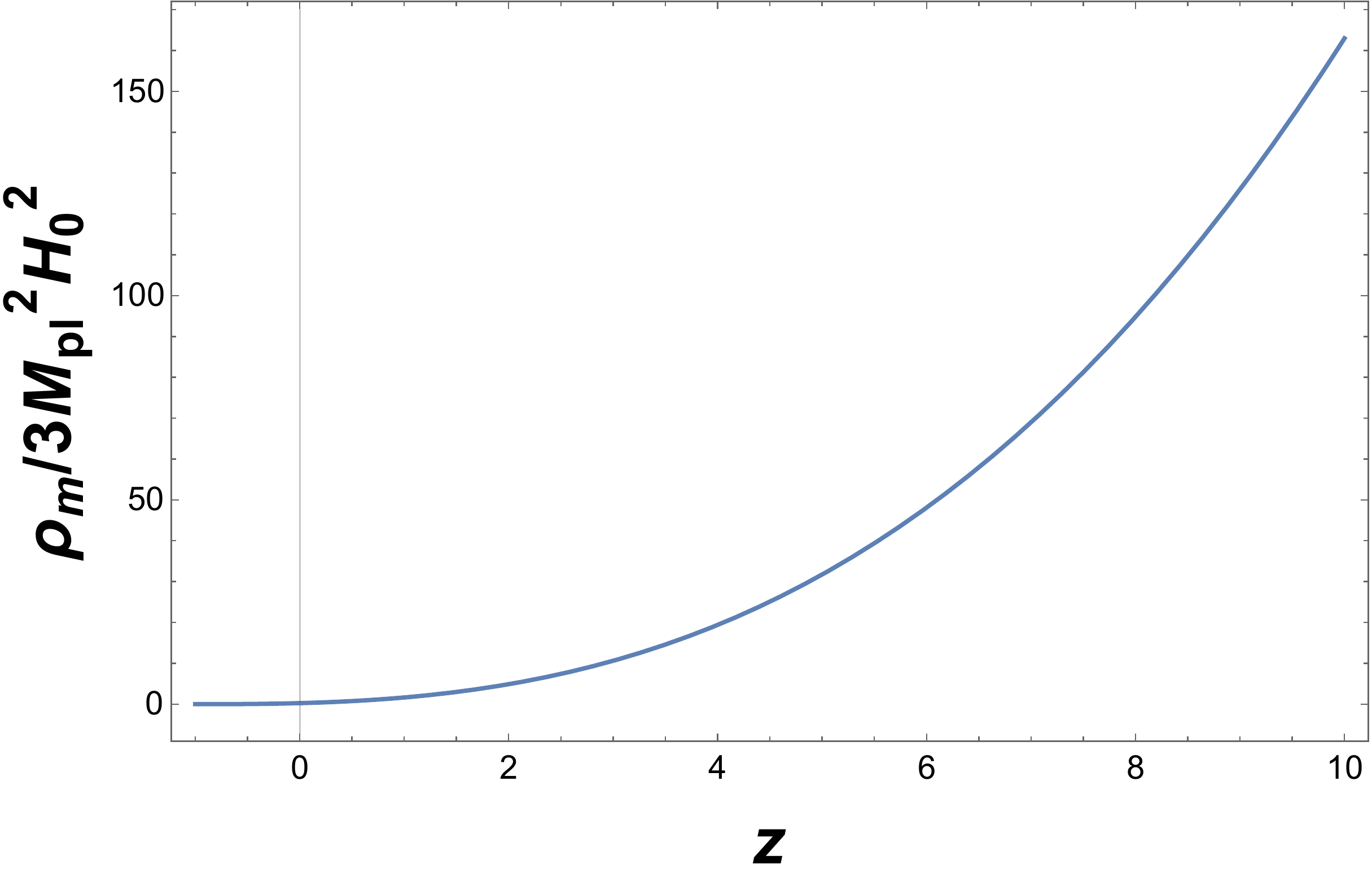} & %
\includegraphics[width=2.2 in, height=2.2 in]{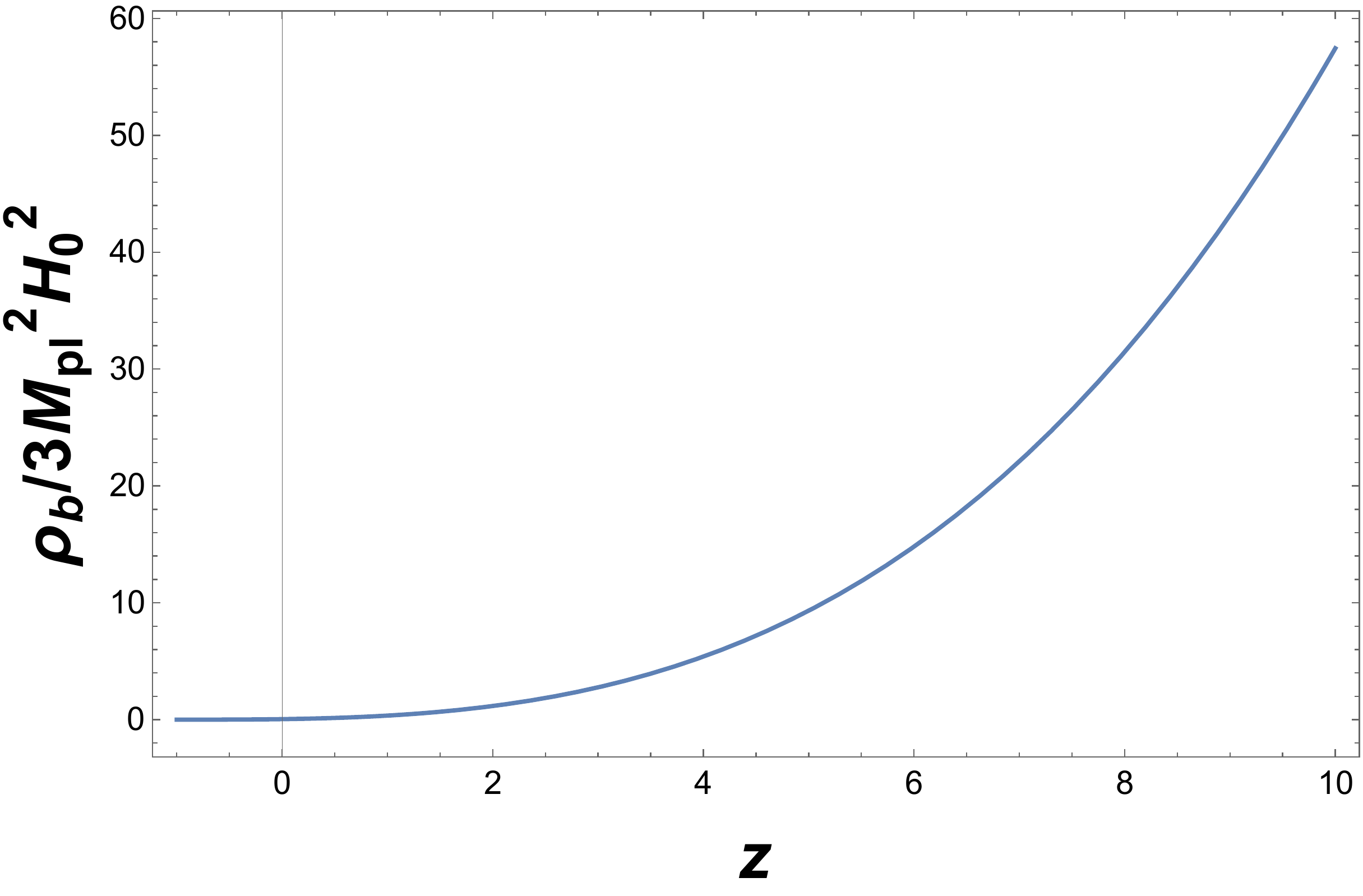} & %
\includegraphics[width=2.2 in, height=2.2 in]{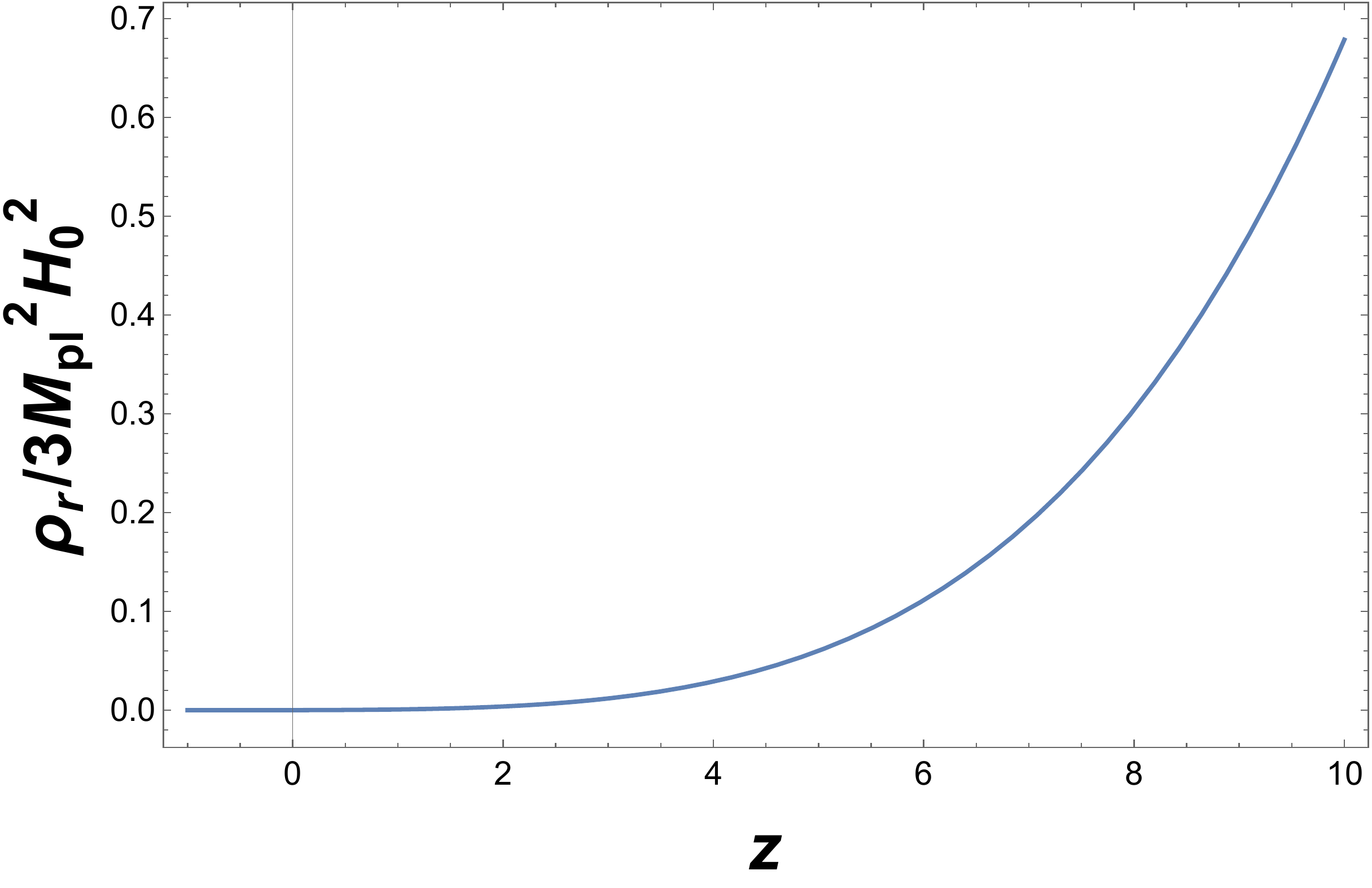} \\ 
\mbox (a) & \mbox (b) & \mbox (c)%
\end{array}%
$%
\end{center}
\caption{{\protect\scriptsize The plots of energy densities $\protect\rho_m$%
, $\protect\rho_b$ and $\protect\rho_r$ for the model. The plots clearly
indicate the faster decrease of the radiation than the baryonic matter and
faster than the dark matter.}}
\end{figure}

\begin{figure}[tbp]
\begin{center}
$%
\begin{array}{c@{\hspace{.1in}}cc}
\includegraphics[width=2.2 in, height=2.2 in]{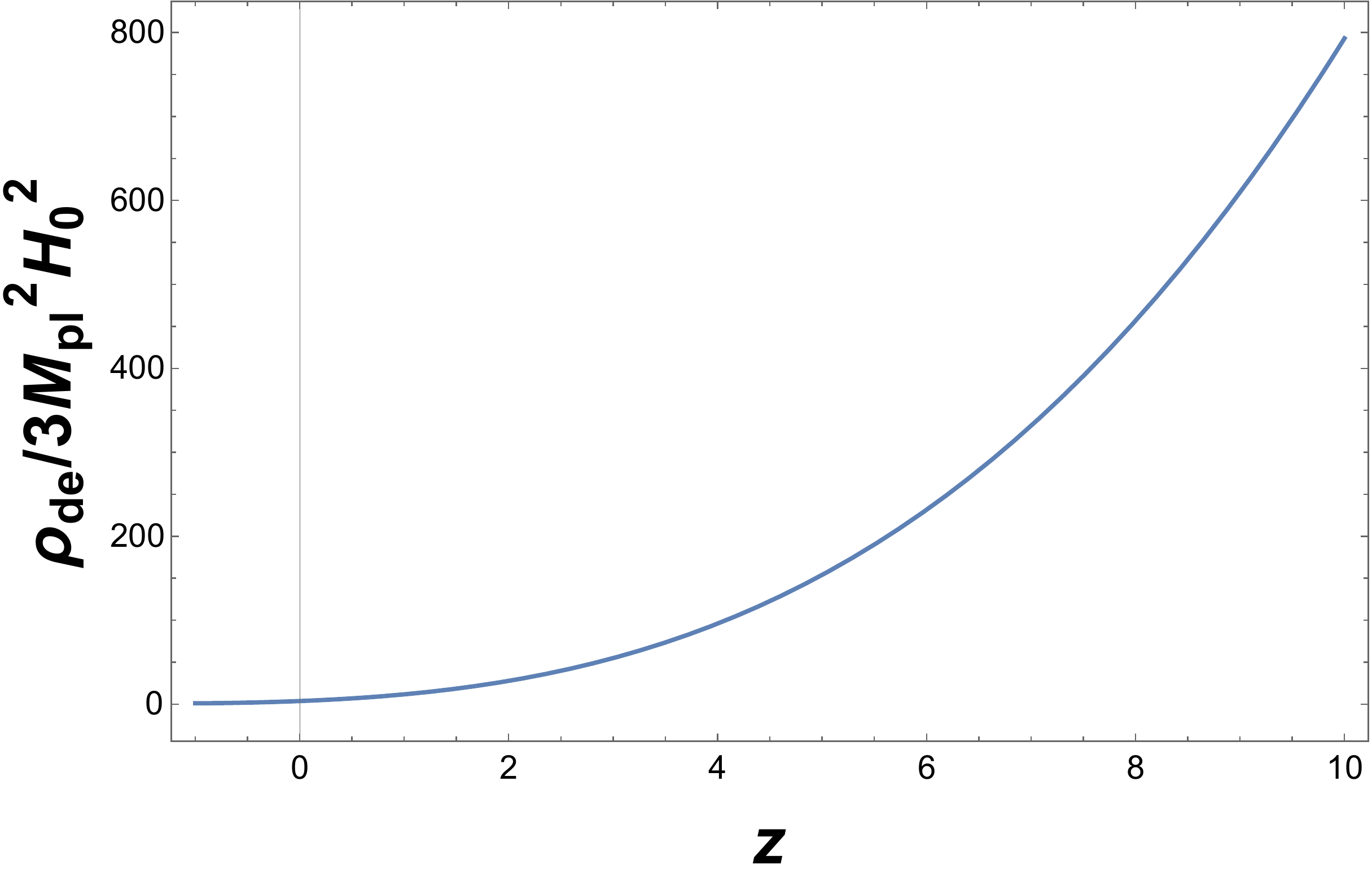} & %
\includegraphics[width=2.2 in, height=2.2 in]{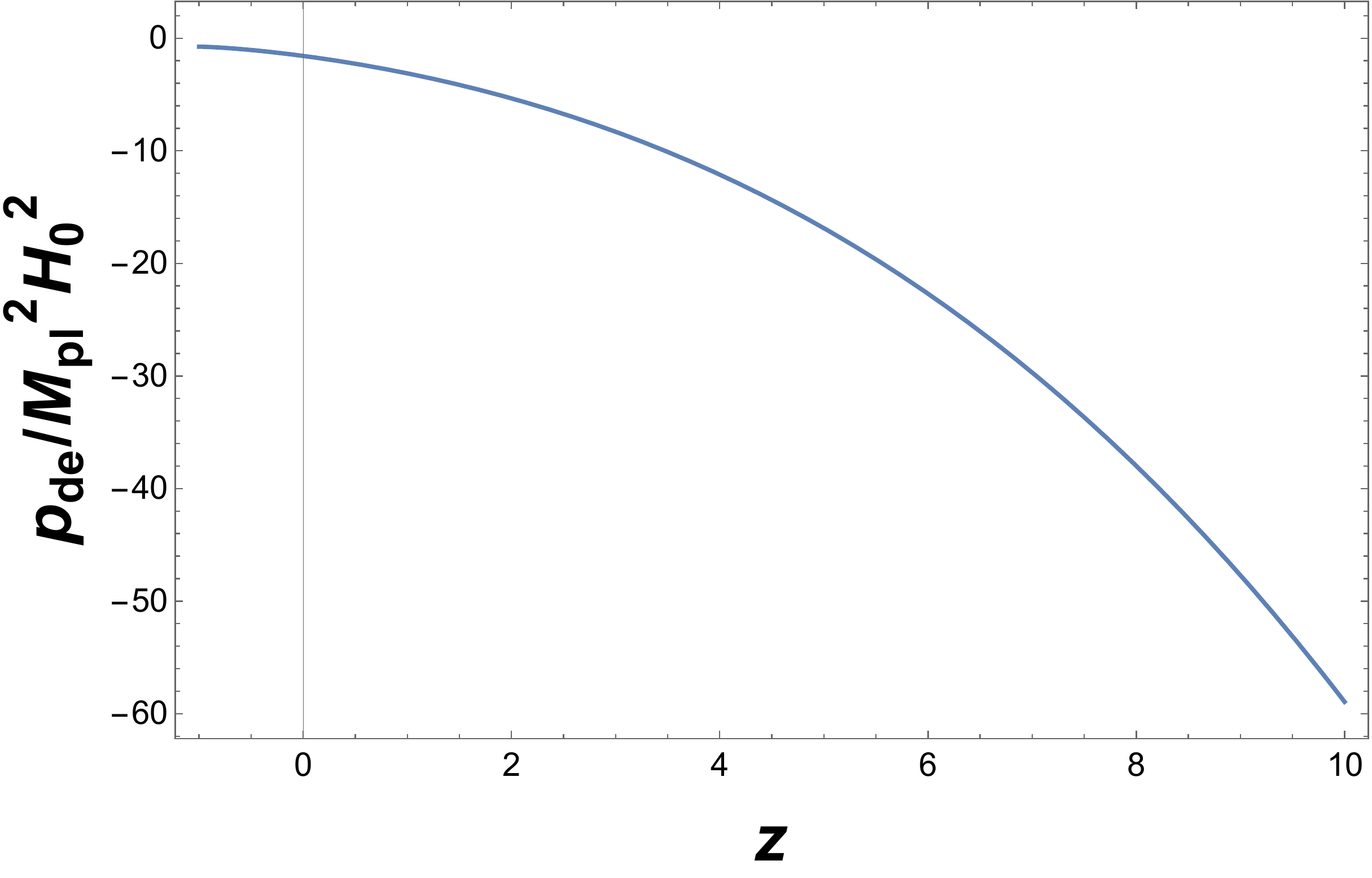} & %
\includegraphics[width=2.2 in, height=2.2 in]{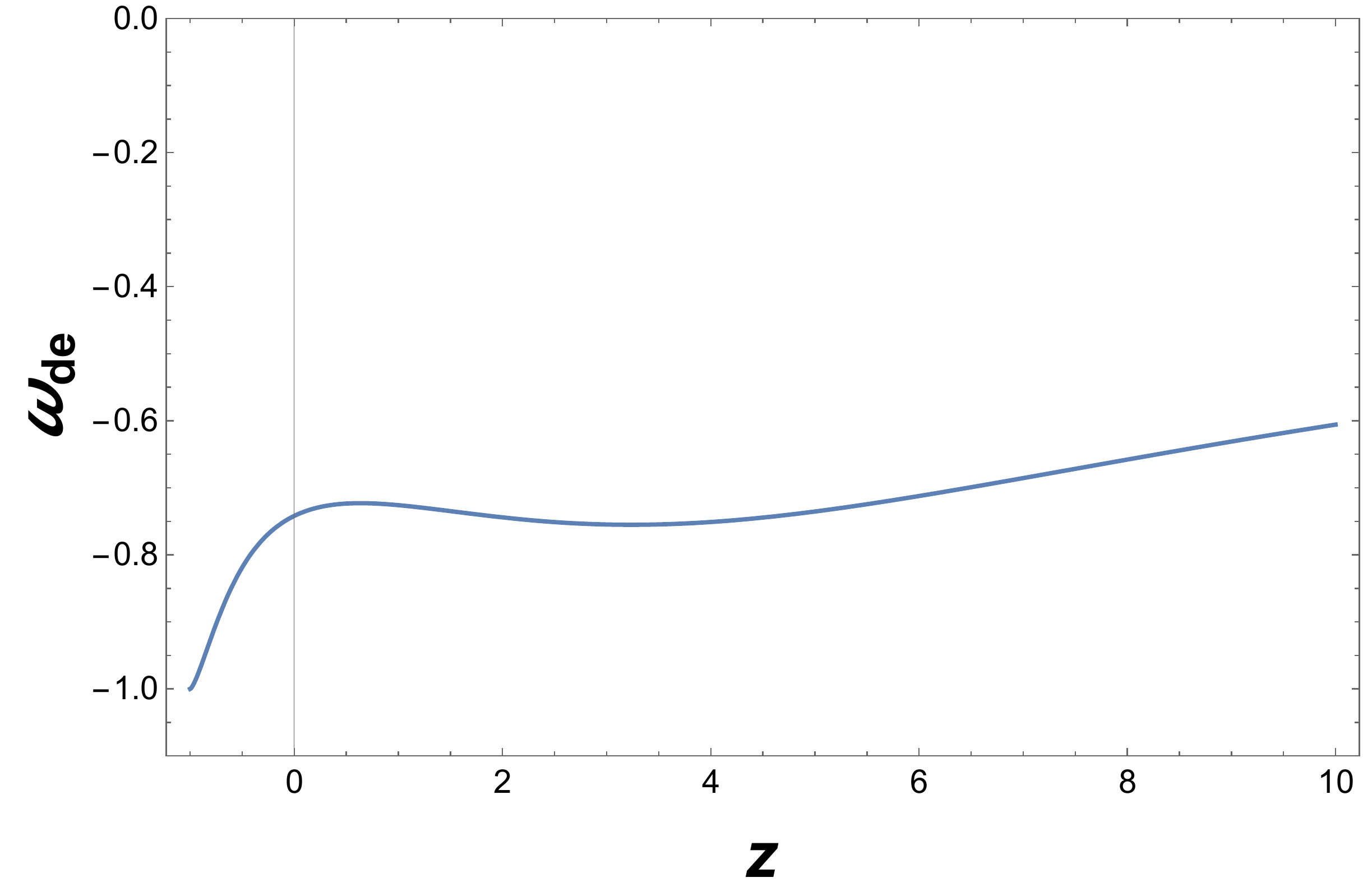} \\ 
\mbox (a) & \mbox (b) & \mbox (c)%
\end{array}%
$%
\end{center}
\caption{{\protect\scriptsize The plots of $\protect\rho _{d}$, $p_{d}$ and $%
\protect\omega _{d}$ for the model.}}
\end{figure}

\section{Energy conditions}

\label{sec4}

There are several prevalent energy conditions (ECs) in general relativity
that set restrictions to prevent some regions having negative energy
density. In other words, ECs are considered a valid generalization of the
energy momentum tensor to the whole Universe, where energy density can never
be negative. The plausibility of numerous significant singularity problems
involving black holes, and wormholes, and many others are extensively
examined using ECs. The energy conditions can mainly be defined in two ways:
(i) geometrically, where ECs are well expressed in terms of Ricci tensor or
Weyl tensor, and (ii) physically, where ECs are expressed as a function of
the physical world. The ECs are defined as follows.

\begin{itemize}
\item Weak energy condition (WEC) $\Leftrightarrow$ $\rho_{total} \geq 0 $, $%
\rho_{total}+p_{total} \geq 0 $,

\item Null energy condition (NEC) $\Leftrightarrow$ $\rho_{total}+p_{total}
\geq 0 $,

\item Strong energy condition (SEC) $\Leftrightarrow$ $%
\rho_{total}+3p_{total} \geq0 $,

\item Dominant energy condition (DEC) $\Leftrightarrow$ $\rho_{total} \geq 0 
$ , $\rho_{total} \geq |p_{total}| $ .
\end{itemize}

Additionally, the energy density ($\rho$) and isotropic pressure ($p$) of
the model can be expressed in terms of potential energy ($V(\phi)$) and
scalar field $\phi$ respectively. Thus the point-wise ECs in GR are defined
as:

\begin{itemize}
\item WEC $\Leftrightarrow$ $V(\phi) \geq \frac{\dot{\phi}^2}{2}$,

\item NEC: $\forall \, V(\phi)$,

\item SEC $\Leftrightarrow$ $V(\phi) \leq \dot{\phi}^2$,

\item DEC $\Leftrightarrow$ $V(\phi) \geq 0.$
\end{itemize}

The examination of various energy conditions to determine whether weak,
null, strong and dominant ECs are satisfied in the given scenarios is
another crucial topic covered here. We plot some statistics for interacting
scenarios and explore various energy conditions for our model.

\begin{figure}[tbp]
\begin{center}
$%
\begin{array}{c@{\hspace{.1in}}cc}
\includegraphics[width=2.2 in, height=2.2 in]{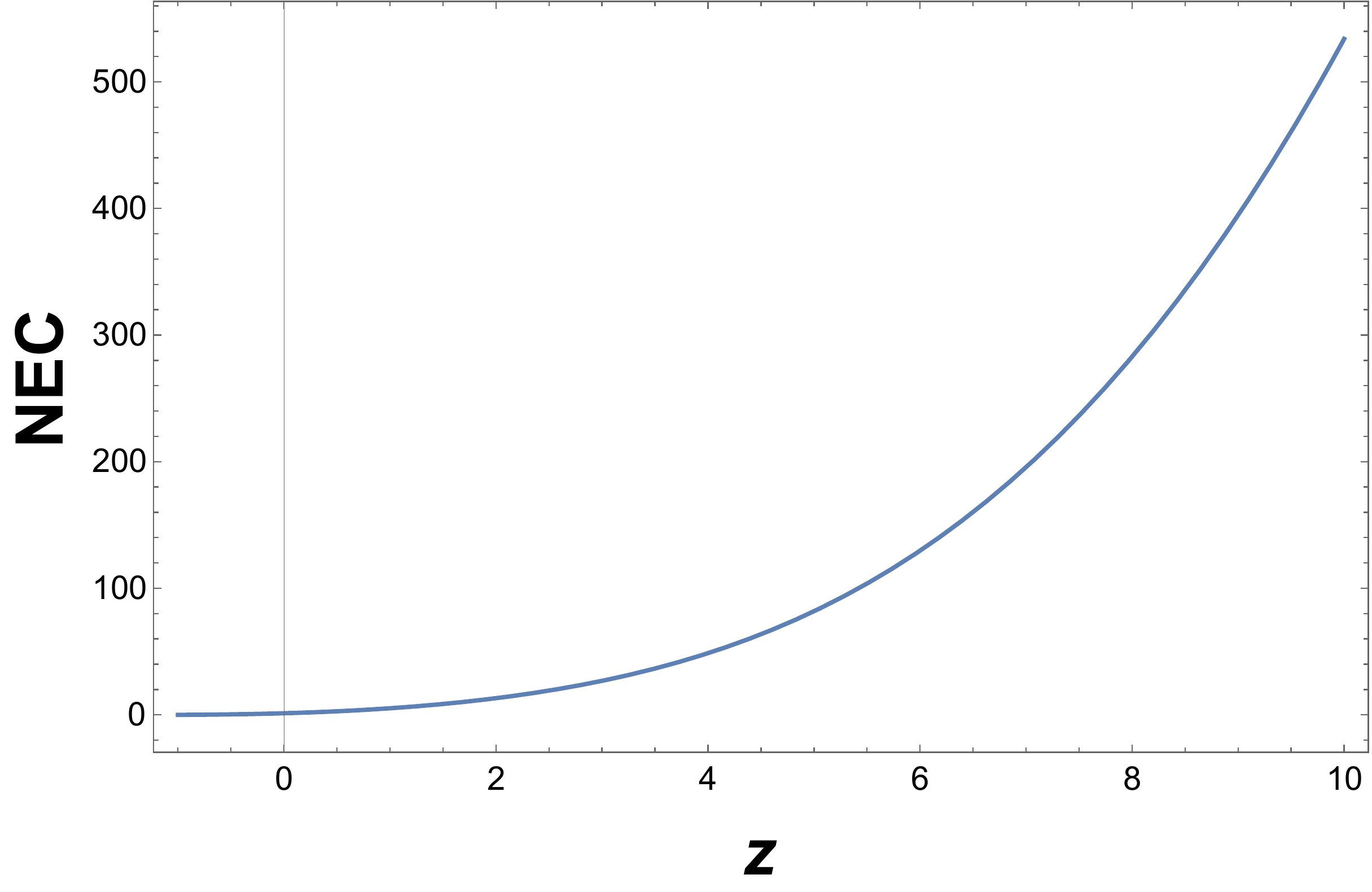} & %
\includegraphics[width=2.2 in, height=2.2 in]{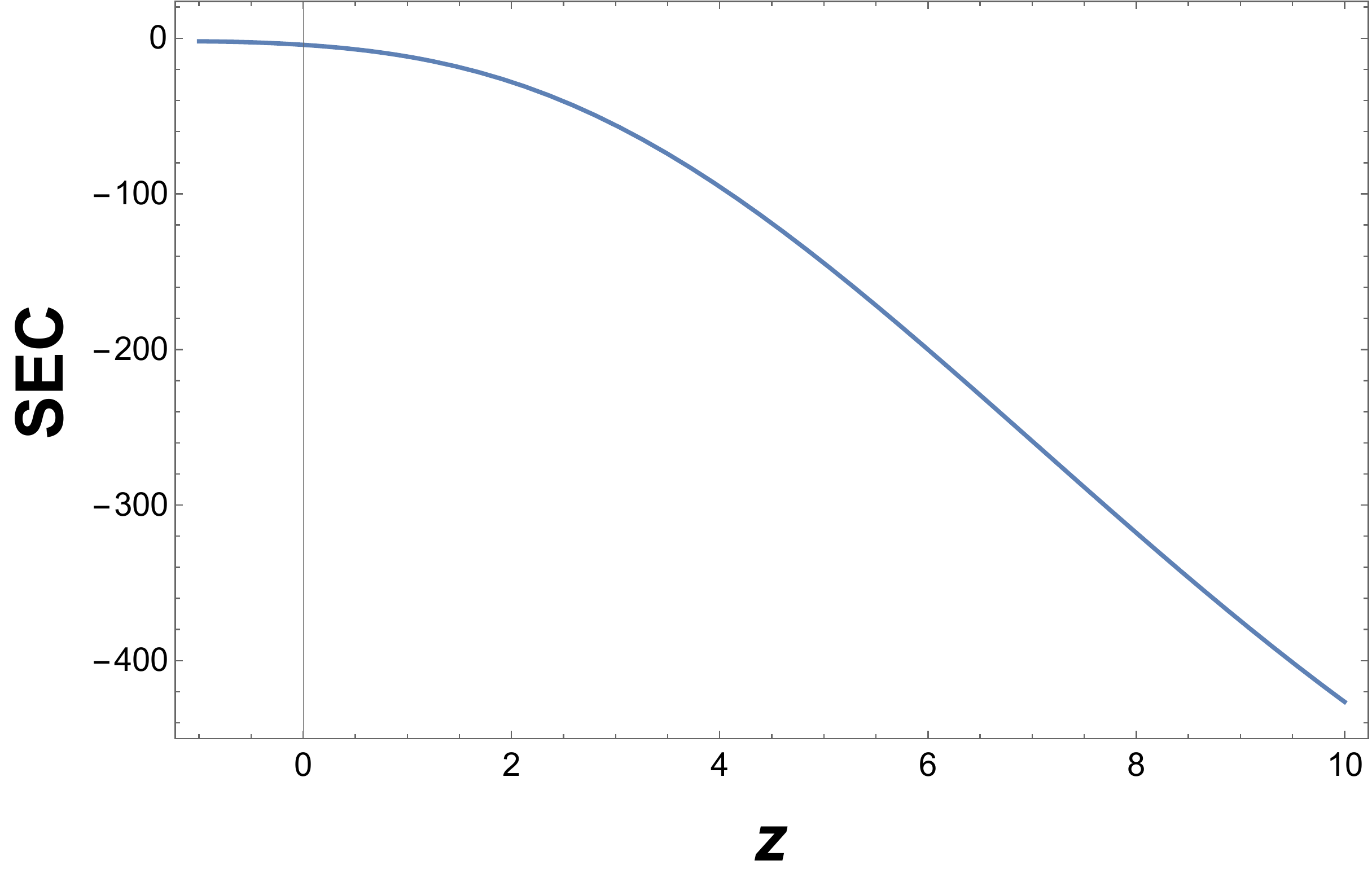} & %
\includegraphics[width=2.2 in, height=2.2 in]{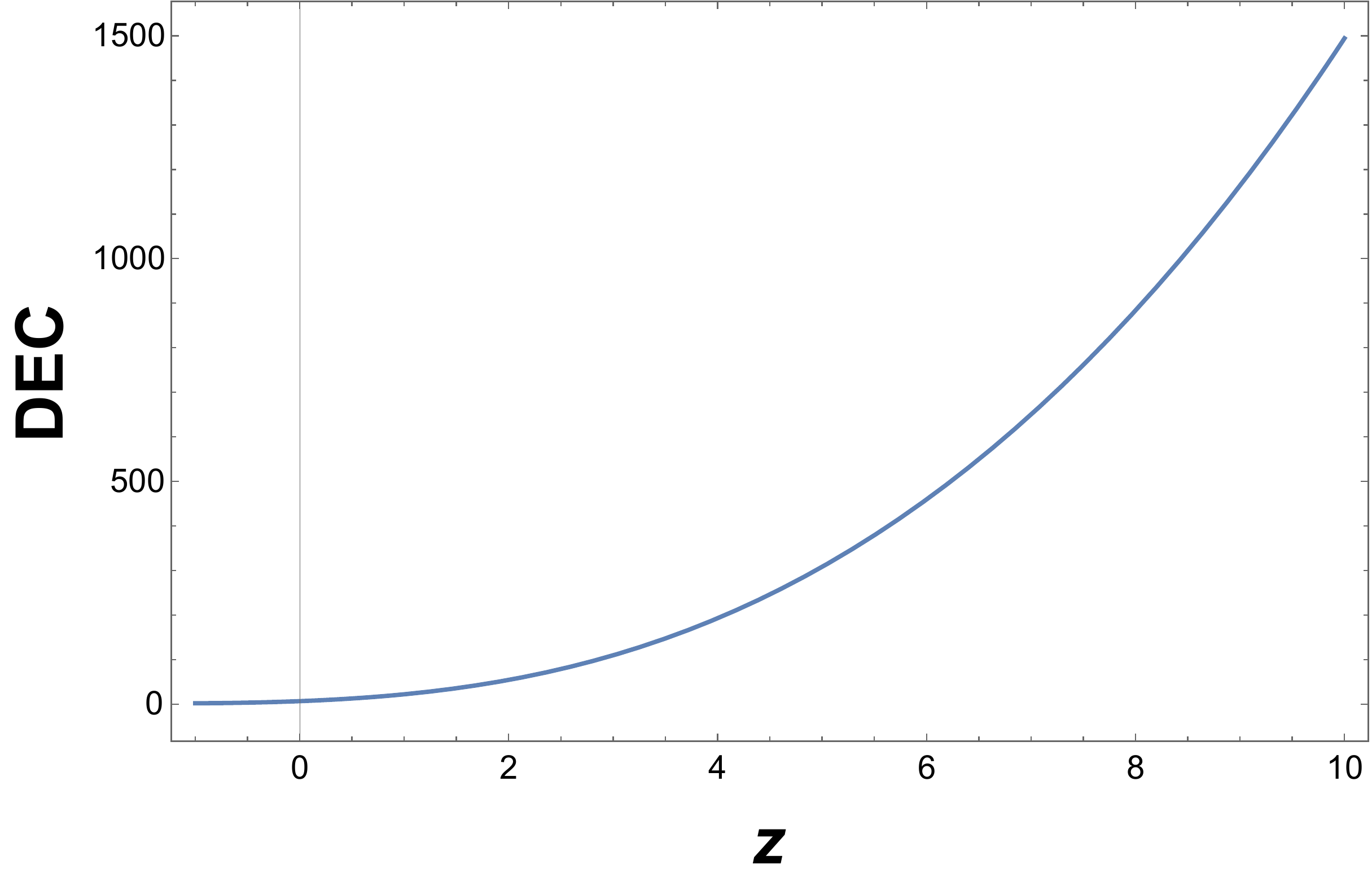} \\ 
\mbox (a) & \mbox (b) & \mbox (c)%
\end{array}%
$%
\end{center}
\caption{{\protect\scriptsize The plots of $Null\ Energy\ Condition\ (NEC)$, 
$Strong\ Energy\ Condition\ (SEC)$ and $Dominant\ Energy\ Condition\ (DEC)$
for the model with suitable units of energy density (}$\protect\rho $%
{\protect\scriptsize ) and pressure (p) as defined (}$H_{0}^{2}$%
{\protect\scriptsize ).}}
\end{figure}

\section{Velocity of sound}

\label{sec5}

As it is widely known that, the stability of linear perturbations is a
significant test for the sustainability of a cosmological model. The need
that the speed of sound ($C_s^2$) be sufficiently less than $1$ to prevent
unintended oscillations in the matter power spectrum imposes a strict
limitation, nevertheless. We, here, plot the graph of velocity of sound ($%
C_s^2$) for our obtained models using suitable choice of the appropriate
parameters, as illustrated in the following figure: 
\begin{figure}[]
\begin{center}
$%
\begin{array}{c@{\hspace{.1in}}cc}
\includegraphics[width=3.2 in, height=3.2 in]{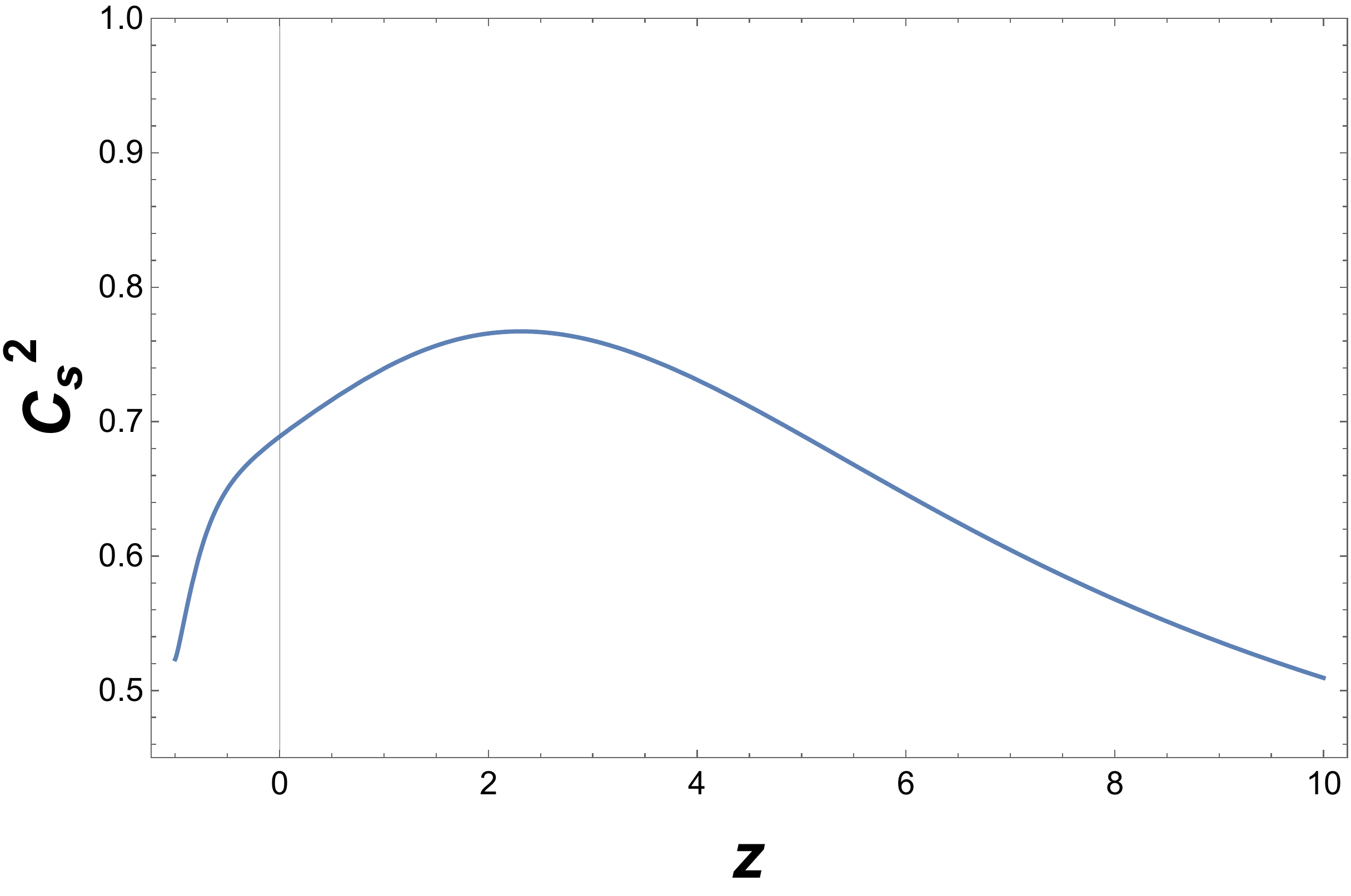} &  &  \\ 
&  & 
\end{array}%
$%
\end{center}
\caption{{\protect\scriptsize The plots of $C_s^2$ for the model.}}
\end{figure}

\section{Swampland conjecture}

\label{sec6}

In the present era, there is a growing interest in linking modern
cosmological models with quantum gravity theory, which can easily
distinguish between the effective field models to the nominal string theory.
In order to choose field potentials properly, Swampland criteria have just
been established. Therefore, in this section, we shall shed light on how the
refined swampland programme conjecture and dark energy are related. The
swampland conjecture, as we are aware, questions a number of structures,
including the physics of black holes, inflation, and other cosmological
ideas. The connection of practical cosmological models with gravity's
quantum theory, which can distinguish between the effective field models
associated to the nominal string landscape, has recently attracted
considerable interest. Swampland criteria were recently established to help
with field potential selection. Excitement was specifically sparked by the
fact that precise de Sitter solutions with positive cosmological constants
do not match with the string landscape, making it impossible to link high
energy arrangements to fundamental theories. \newline
The swampland criteria of the string theory bound with the quintessence
model of DE were recently derived \cite{vafa}, \cite{ooguri}. Numerous works
followed the work in \cite{vafa}, \cite{ooguri} approaching the Swampland
criterion in the connection of DE's quintessence models, for instance \cite%
{swam1} - \cite{swam12}. The validity of the $f(R)$ theory was recently
studied by the researcher in relation to the Swampland conjecture \cite%
{swam1}, \cite{swamO1}. When a field theory contradicts quantum gravity, the
Swampland requirements constrain and regulate it \cite{swam13}. Although
these hypotheses are still in their infancy and address some issues, they
are rapidly evolving and changing in some situations like warm inflation 
\cite{warm}. We therefore choose to make a connection between these
hypotheses and dark energy. In subsequent works, we'll endeavour to develop
this concept and make a more substantial connection between the swampland
programme, dark energy, dark matter, and in action. Researchers take into
account several models for dark energy, and studied their cosmological
applications, but the nature of the concept is still unclear. \newline
Here, we investigate the cosmological restrictions put forth by two string
Swampland criteria. These criteria include a lower restriction on $\frac{%
\mid \triangledown _{\phi }V\mid }{V}$ when $V>0$ and an upper bound on the
range that scalar fields can travel. The major disadvantage of these two
criteria seems that they generally in conflict with inflationary theories.
We discover that certain quintessence models can meet these restrictions at
the same time when we apply the same criterion to dark energy in the current
epoch. We contend that the universe will experience a phase change within a
few Hubble times if the two Swampland requirements are true. These
conditions intensify the drive for upcoming dark energy equation of state $%
\omega $ measurements and equivalence principle tests for dark matter. With
this motivation, we propose, the refined swampland conjecture in the below
form.\newline
Conjecture 1. $\frac{|V^{\prime }|}{V}>C_{1}M_{pl}$ \newline
Conjecture 2. $\frac{|V^{\prime \prime }|}{V}<-C_{2}M_{pl}^{2}$

As we have already noted that the notion to forecast the acceleration in the
Universe is to fill it with an exotic type of matter that satisfies the
inequality $1+3\omega <0$. The energy that causes the acceleration,
according to the measurements, satisfies $\omega \simeq -1$. The development
of the EoS parameter $\omega $ for our model was already covered in earlier
sections, thus to get an appropriate matter field that produces unusual
behavior and is capable of exhibiting repulsive effects whose cause is given
by dark energy (scalar field) we will consider the following equations,

\begin{equation}  \label{27}
\rho=\frac{1}{2}\dot{\phi}^2+V(\phi),
\end{equation}

\begin{equation}  \label{28}
p=\frac{1}{2}\dot{\phi}^2-V(\phi).
\end{equation}

where the terms $\frac{1}{2}\dot{\phi}^2 $ and $V(\phi) $ refer to the
kinetic energy $(KE) $ and potential energy $(PE) $ of the scalar field. So
the term $\omega = \omega(t) $ \textit{i.e.} can no more be treated as a
constant. The quintessence or phantom model is consistent with the
observations provided $\omega\simeq -1 $. Thus, we need $\dot{\phi}%
^{2}<<V(\phi) $ \textit{i.e.} the $KE$ of $\phi$ is insignificant in
comparison to the $PE$. In this study, we consider that $\phi$ is the only
source of DE with $V(\phi )$, so one can consider energy density and
pressure of scalar field as $\rho _{\phi}$ and $p_{{\phi}}$ respectively for
flat FLRW space-time under Barrow's scheme \cite{barrow} using Eqs. (\ref{27}%
) and (\ref{28}) as 
\begin{equation}  \label{29}
\rho=\frac{1}{2}\dot{\phi}^2+V(\phi)=\rho_{\phi},
\end{equation}
\begin{equation}  \label{30}
p=\frac{1}{2}\dot{\phi}^2-V(\phi)=p_{\phi}.
\end{equation}
The $KE$ and $PE$ can be obtained by solving the Eqs. (\ref{29}) and (\ref%
{30}). Fig. 4 demonstrates the potential energy $V(\phi) $ plots \textit{%
w.r.t.} scalar field $\phi$ for the same considered values of model
parameters as we have taken in Fig. 1, 2. From Fig. 4, we notice that the
potential $V(\phi) $ is present in the interval $-1<\phi<0 $ and $%
V(\phi)\simeq0 $ at $\phi\simeq0 $. Therefore, we can predict that the
scalar field $\phi $ is the only source of DE with potential $V(\phi) $.
Thus we conclude that our model is an accelerating dark energy model. 
\newline

To understand the nature of dark energy, we have considered its only source
of energy as scalar field $\phi $, which plays the role of quintessence
model, therefore in this regard we have analyzed our model in the connection
with Swampland conjecture. Conjecture 1 of this criteria is linked with
scalar field $\phi $, whereas conjecture 2 is connected with scalar
potential $V$. We have composed the behavior of scalar field with respect to
redshift for our interacting model in the figure below. \newline


\begin{figure}[]
\begin{center}
$%
\begin{array}{c@{\hspace{.1in}}cc}
\includegraphics[width=2.2 in, height=2.2 in]{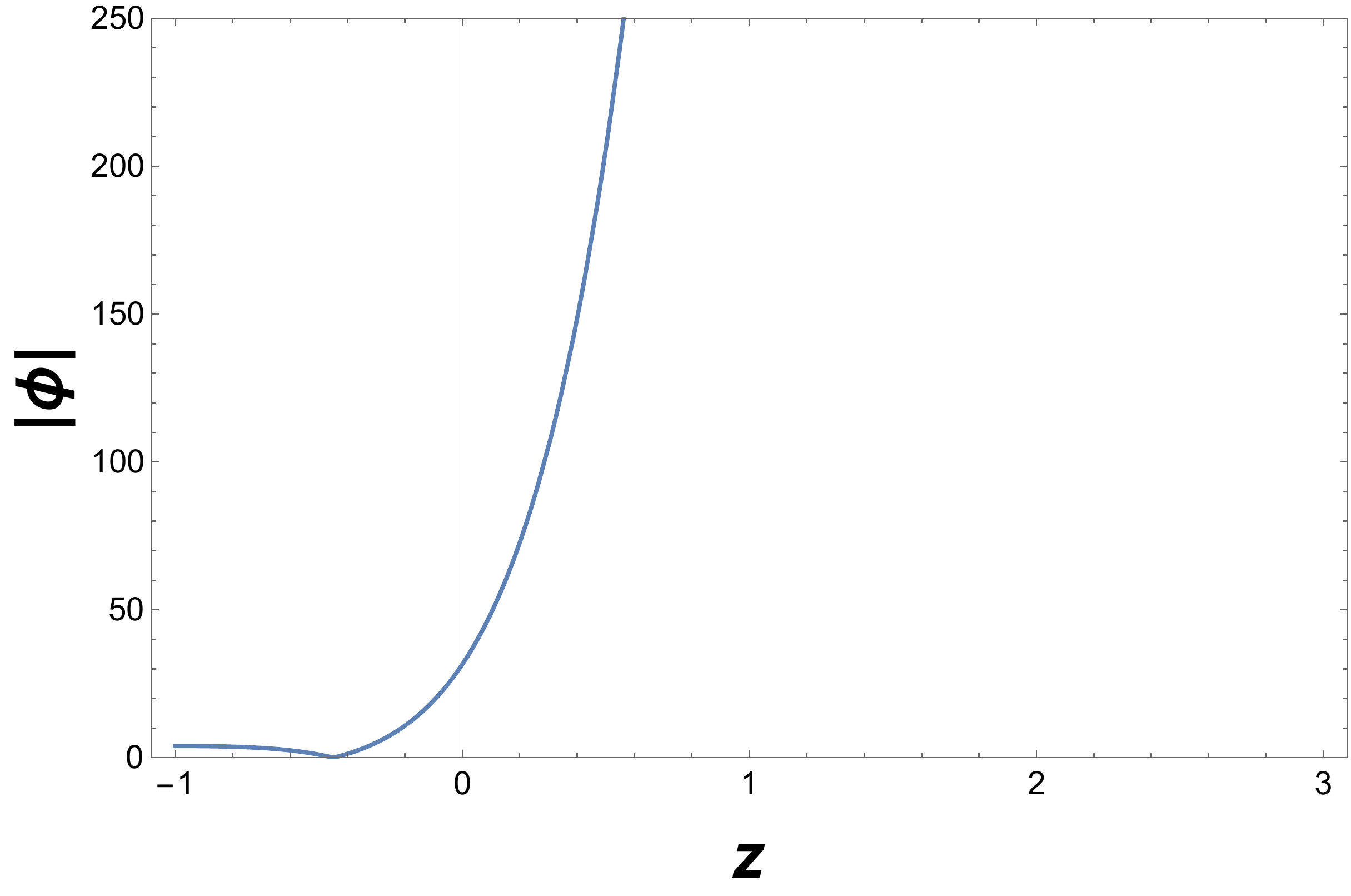} & %
\includegraphics[width=2.2 in, height=2.2 in]{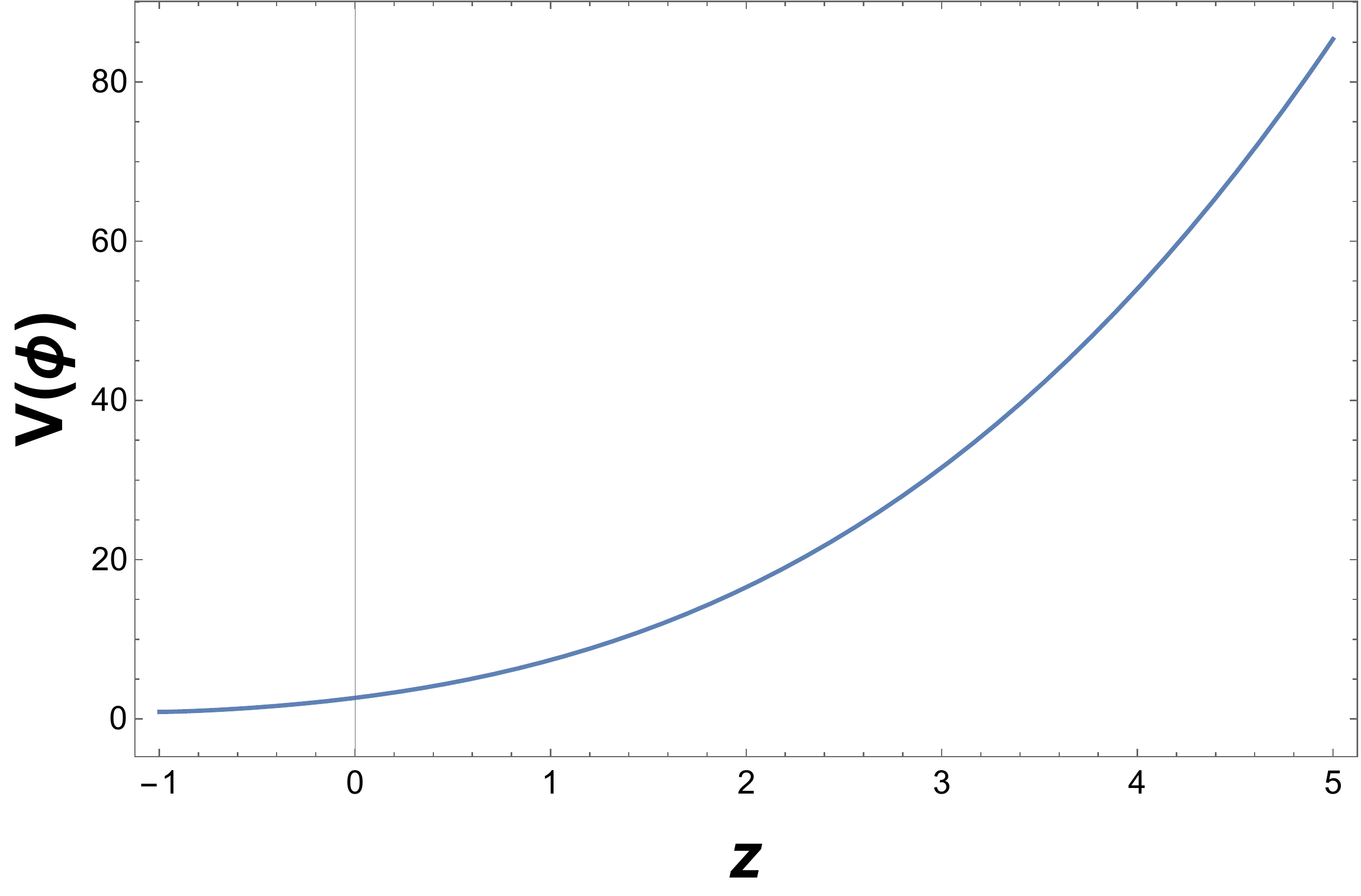} & %
\includegraphics[width=2.2 in, height=2.2 in]{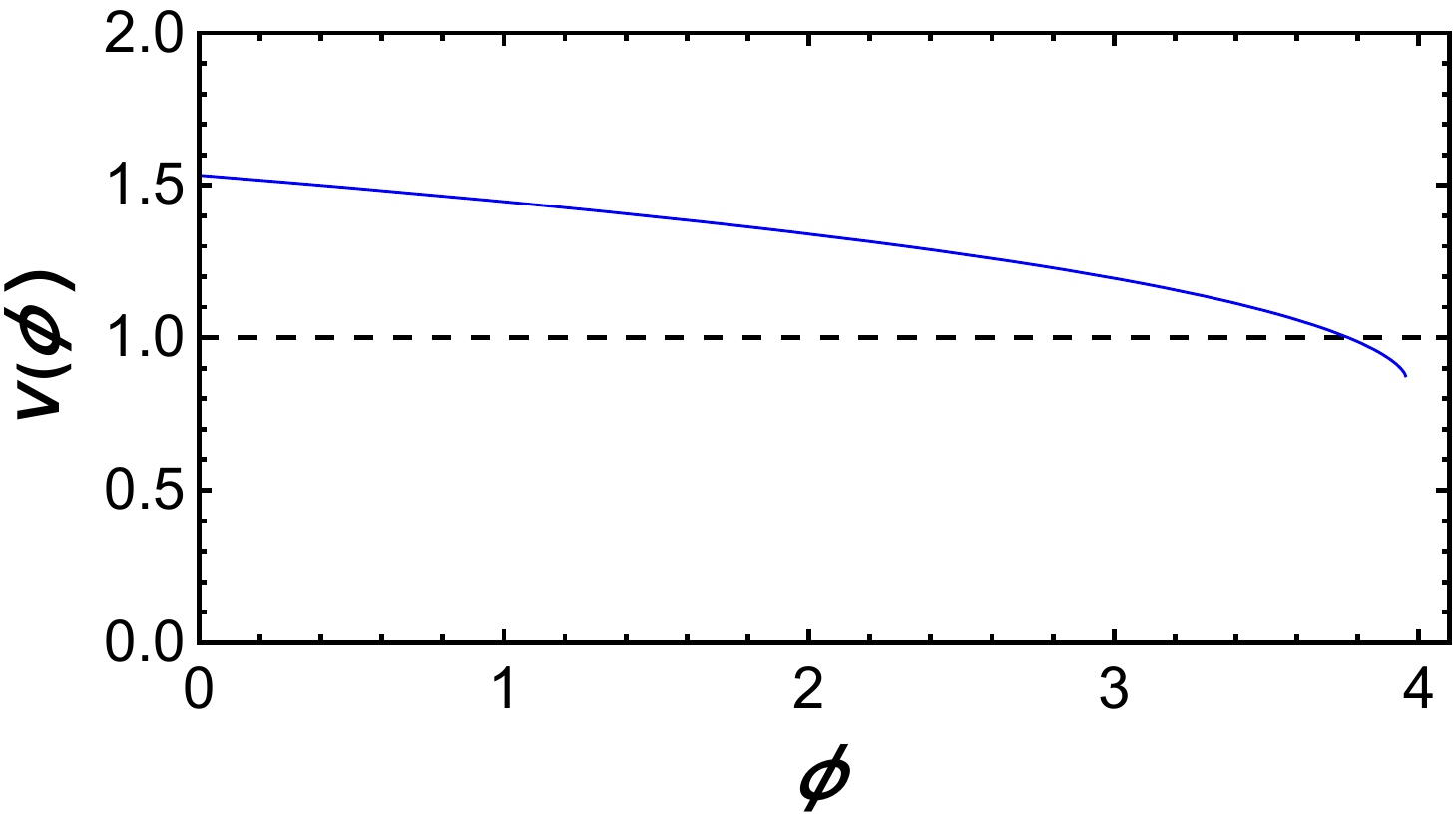} \\ 
\mbox (a) & \mbox (b) & \mbox (c)%
\end{array}
$%
\end{center}
\caption{{\protect\scriptsize The plots of scalar field $\protect\phi$,
potential $V(\protect\phi)$ w.r.t. $z$ and scalar filed correspondence $V( 
\protect\phi)$ vs $\protect\phi$ for the model.}}
\end{figure}

The below figure illustrates the behavior of refined swampland conjecture's
components plotted for two distinct potentials connected to an interacting
scenario for various parameters of the universe. We are aware that each of
the swampland conjectures in the literature has positive values of unit
order and the component $C_{2}$, is smaller than component $C_{1}$. The
interacting model demonstrates satisfaction of both conjectures, as shown in
figure. The Swampland criterion, which has been proposed for a consistent
theory of gravity in this regard and which satisfies the requirement $\frac{%
|V^{\prime }|}{V}>C_{1}\approx O(1)$, demonstrates that this dark energy
model accords better with the Swampland criteria. 
\begin{figure}[tbp]
\begin{center}
$%
\begin{array}{c@{\hspace{.1in}}cc}
\includegraphics[width=2.2 in, height=2.2 in]{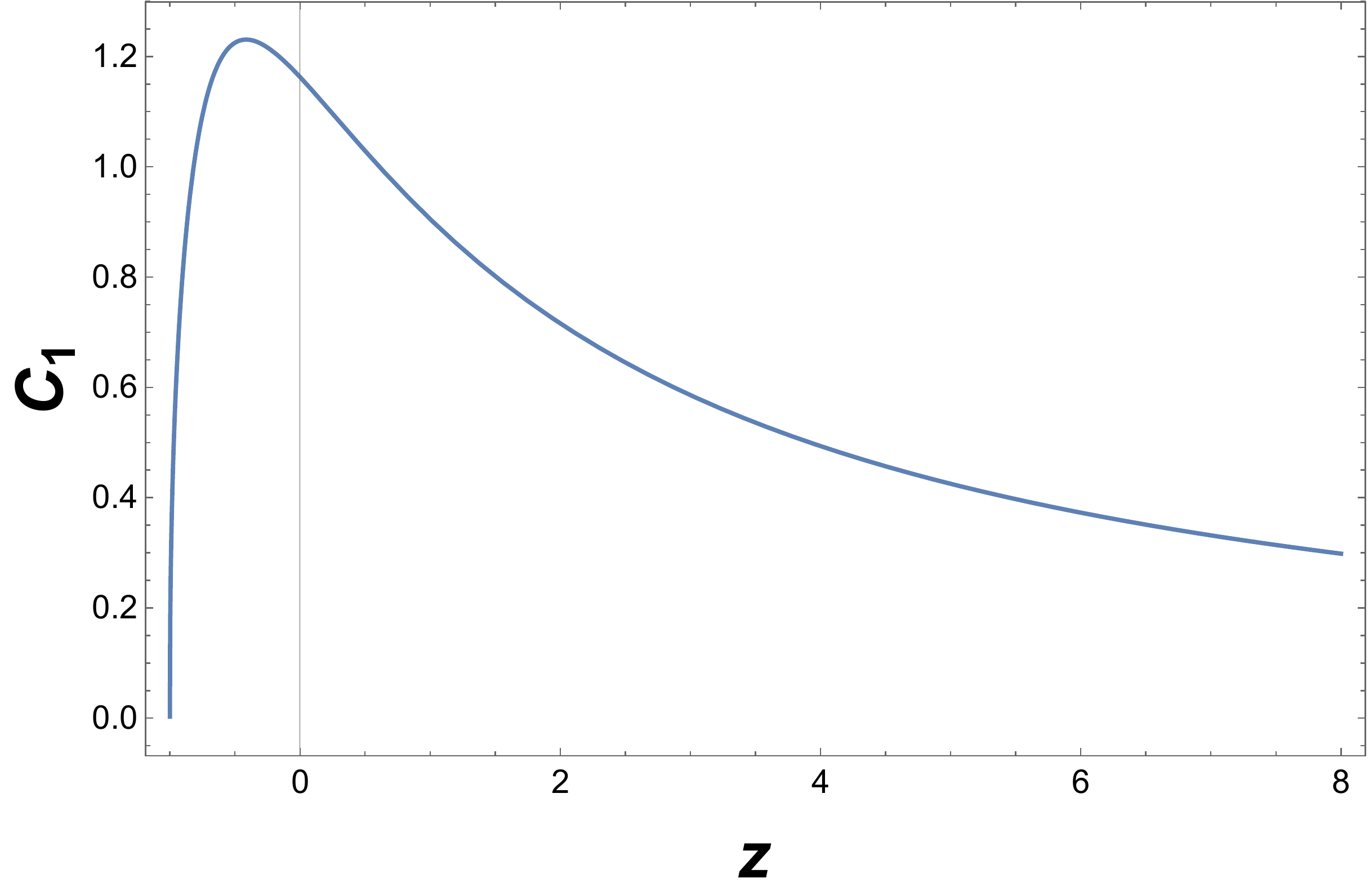} & %
\includegraphics[width=2.2 in, height=2.2 in]{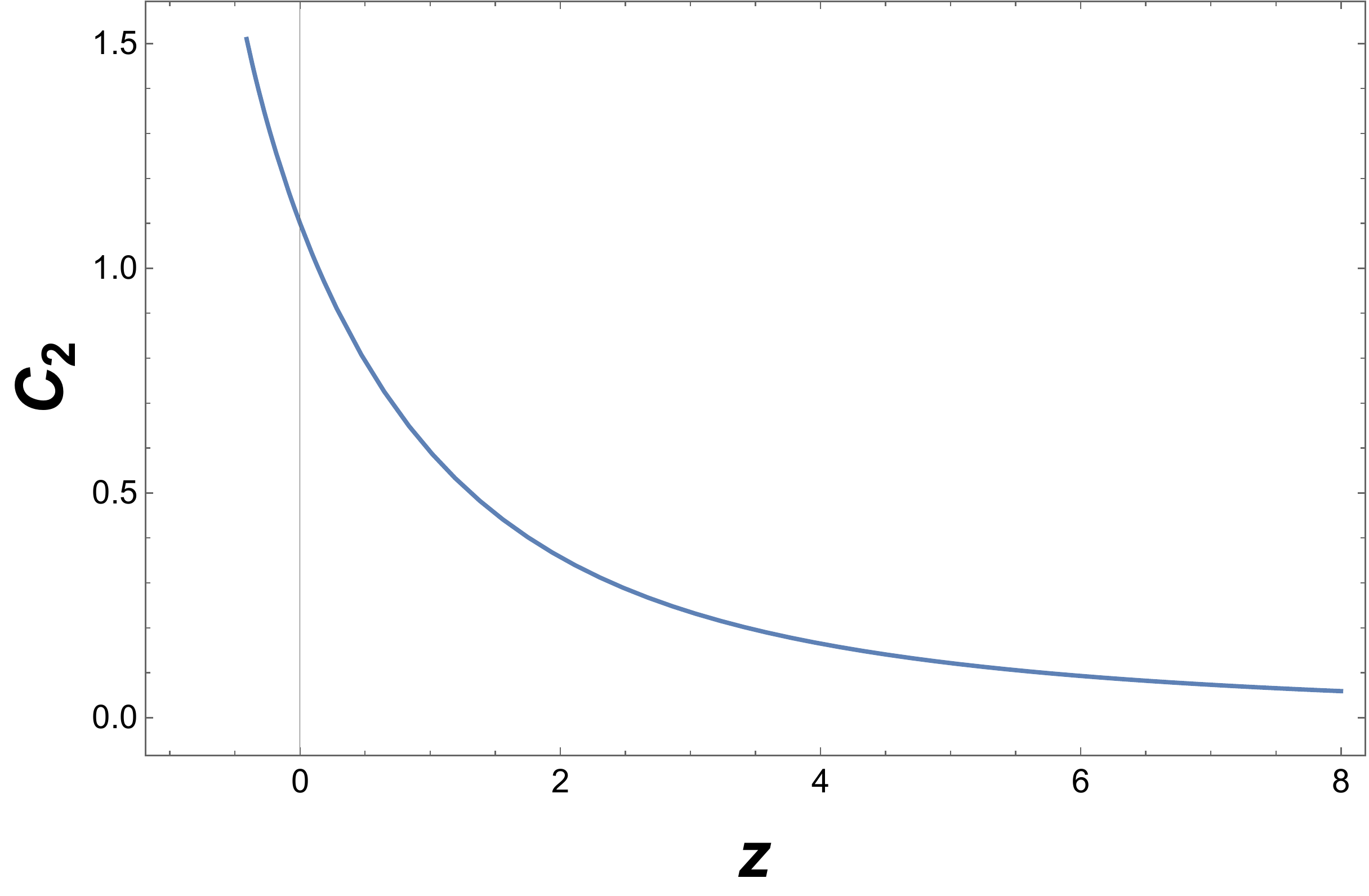} &  \\ 
&  & 
\end{array}%
$%
\end{center}
\caption{{\protect\scriptsize The plots of swampland conjecture 1 and 2 for
the model.}}
\end{figure}



\section{Conclusion}

\label{sec7}

We have examined an interacting dark energy cosmological model in the
classical theory of gravity. The interacting scenario discussed here
provides an insight into the evolution of cosmological parameters. In order
to find a consistent solution, we have considered a parametrization of
Hubble parameter, which provide a smooth transition from early deceleration
to present acceleration and discuss the late evolution of the Universe. In
order to discuss the Universe's evolution in particular the late-time
behavior, all the concerned cosmological parameters were written in terms of
redshift $z$ after establishing the $(t-z)$ relationship. All the
cosmological parameters concerned with one model parameter $n$ only that
could be constrained from any cosmological datasets but we have used the
constrained value of $n=1.43$ as found in our earlier works \cite{RNSKJP}, 
\cite{SMSBSKJP} for our analysis here. The evolutionary profiles of energy
densities of radiation, baryonic \& dark matter, ark energy, pressure and
equation of state parameter are shown in figures 1 and 2. We can observe the
faster decrease of the radiation energy than the baryonic matter, which is
faster than the dark matter. Panel 2(b) highlights the profile of pressure
of dark energy which is more negative in the past, increases in a concave
upward way and finally tends to zero in the future. The negative pressure of
dark energy in due course of evolution favours the standard lore. Figure
2(c) shows the redshift evolution of equation of state parameter, which is
approaching to $-1$ in the far future assuming the negative values in the
expected ranges around $z=0$. We have discussed different energy conditions
for interacting models and shown them with graphical representations and
also we have discussed the velocity of sound for our obtained model using
suitable choice of appropriate parameters and is shown in Figure 4. Further
we reconstructed the potential of the scalar field and challenged the
refined swampland conjecture. The imprints on the growth of matter as well
as dark energy fluctuations in interacting models of dark matter - dark
energy have been studied in several papers \cite{pert1}, \cite{pert2}, \cite%
{pert3} at perturbative level. For example, in \cite{pert1}, the
perturbation equations are numerically solved, and it is discovered that at
the early stages of development, visible imprints on the fluctuations in
dark energy density are seen for larger rates of interaction strength for
the coupling component. The evolution of cosmological perturbations is
investigated in a model with dynamical dark energy in \cite{pert4}, which
interacts with dark matter non-gravitationally. The influence of the
strength of the additional interaction between the dark components on the
evolution of density and velocity perturbations in them is analyzed for
quintessence and phantom types of dark energy. We may expect the similar
kinds of result with as the similar coupling term considered here and the
detailed study is deferred to our future investigation.

{\LARGE Acknowledgement}

SKJP thank IUCAA, Pune for hospitality, where a large part of this paper was
written. The work of KB was supported in part by the JSPS KAKENHI Grant
Number JP21K03547.

\textbf{Data Availability Statement:}{\ No Data associated in the manuscript}

\end{document}